\documentclass[twocolumn,superscriptaddress,preprintnumbers,amsmath,amssymb,prl]{revtex4}
\usepackage{graphicx}

\begin{document}
\newcommand{\ri}{{\rm i}}
\newcommand{\il}{{{\rm i}\xi_l}}
\newcommand{\kk}{{({\rm i}\xi_l,k)}}
\newcommand{\eps}{{\varepsilon}}

\thispagestyle{empty}

\title{Effect of increased stability of peptide-based coatings in the Casimir regime
via nanoparticle doping
}

\author{
G.~L.~Klimchitskaya}
\affiliation{Central Astronomical Observatory at Pulkovo of the
Russian Academy of Sciences, Saint Petersburg,
196140, Russia}
\affiliation{Institute of Physics, Nanotechnology and
Telecommunications, Peter the Great Saint Petersburg
Polytechnic University, Saint Petersburg, 195251, Russia}

\author{
V.~M.~Mostepanenko}
\affiliation{Central Astronomical Observatory at Pulkovo of the
Russian Academy of Sciences, Saint Petersburg,
196140, Russia}
\affiliation{Institute of Physics, Nanotechnology and
Telecommunications, Peter the Great Saint Petersburg
Polytechnic University, Saint Petersburg, 195251, Russia}
\affiliation{Kazan Federal University, Kazan, 420008, Russia}

\author{
E.~N.~Velichko}
\affiliation{Institute of Physics, Nanotechnology and
Telecommunications, Peter the Great Saint Petersburg
Polytechnic University, Saint Petersburg, 195251, Russia}

\begin{abstract}
We find that thin peptide films and coatings doped with metallic
nanoparticles are more stable due to the role of electromagnetic fluctuations.
It is shown that for the doped freestanding in vacuum peptide film the Casimir
attraction becomes larger in magnitude. For dielectric substrates coated with
peptide films, the nanoparticle doping leads to a wider range of film
thicknesses where the Casimir pressure is attractive and to larger pressure
magnitudes at the points of extremum. The doping of peptide coatings with
magnetic nanoparticles preserves all the advantages of nonmagnetic ones and
simultaneously imparts superparamagnetic properties to the coating which
could extend significantly the application areas of bioelectronics.
\end{abstract}

\maketitle

It is common knowledge that thin films and coatings based on peptides,
proteins and other biological polimers find increasing bioelectronic and
biomedical applications \cite{1,2}. Peptides are the relatively short chains
of amino acids linked by peptide bonds. These are nonmetallic materials
which posses some electrical conductivity and enjoy wide use in optical
and electronic devices, as well as in biomedical technologies, for creating
thin film transistors, biomarkers, sensors, and biocompatible electrodes
alternative to conventional devices based on the silicon technologies
\cite{3,4,5,6,7,8,9}.

The most general requirements imposed upon peptide-based coatings are the
stability, even in adverse conditions, reproducible diagnostic results,
the ease of fabrication and, in medical applications, biocompatibility and
non-toxicity \cite{10,11,12}. In the last two decades, great progress has
been made in developing miniature bioelectronic devices satisfying these
requirements (see, for instance, Refs. \cite{13,14,15,16,17}). In doing so,
with decreasing the characteristic device dimensions to below a micrometer,
the role of quantum effects and, specifically, the zero-point and thermal
fluctuations of the electromagnetic field, increases in importance.

The fluctuation phenomena and their role in nanoscale science are
the long-explored areas \cite{18,19}. Much experimental and theoretical
attention has been paid to investigation of the
fluctuation-induced van der Waals and Casimir
forces between the boundary surfaces made of inorganic materials
(see, e.g., Refs.~\cite{20,21,22,23} for a review).
It was demonstrated that at separations
below a micrometer these forces may exceed in magnitude
the characteristic electric force and can be used as an actuator in micro-
and nanoelectromechanical devices of the next generations.

These experimental and theoretical advances are largely based on the
fundamental Lifshitz theory which allows calculation of the van der Waals
and Casimir forces between surfaces with known frequency-dependent
dielectric permittivities \cite{20,21,22,23,24}. The Lifshitz theory was
also used to calculate the fluctuation-induced forces acting between
varied in composition organic films \cite{25,26,27,28} and the free
energies of both freestanding in vacuum and deposited on substrates
peptide films \cite{29,30}. It was shown that the free energy of
peptide coating is a nonmonotonous function of the film thickness and
may change its sign. However, the fluctuation-induced pressure in
peptide films and coatings was not considered so far.

In this Rapid Communication, we investigate the pressure in peptide
films and coatings
in the Casimir regime (i.e., for film thicknesses below a micrometer),
and find the effect of increased stability which arises via the
nanoparticle doping. For this purpose, we calculate the Casimir pressure
in the framework of the Lifshitz theory for the ordinary and doped with
metallic nanoparticles peptide films and coatings. Both cases of
nonmagnetic and magnetic nanoparticles are considered. In the latter
case, the coating becomes superparamagnetic which is beneficial for
various applications (the possibility for fabrication of peptide films
with well distributed metallic nanoparticles was demonstrated in Ref.
\cite{31}). According to our results, the doping of a peptide coating
with metallic nanoparticles leads to a larger in magnitude (negative)
minimum value of the Casimir pressure which makes the coating more
stable. Taking into consideration that stability is an essential
feature required of peptide coatings, future prospects for the use of
this effect are discussed.

We consider the three-layer system consisting of a vacuum and a doped peptide
film of thickness $a$ deposited on a nonmagnetic dielectric substrate.
Separate layers of this system are described by the dielectric permittivities
$\eps^{(0)}=1$, $\eps^{(1)}(\omega)$, $\eps^{(2)}(\omega)$ and
magnetic permeabilities $\mu^{(0)}=1$, $\mu^{(1)}(\omega)$ and $\mu^{(2)}=1$,
respectively. The substrate is assumed to be thicker than $2~\mu$m in which
case it can be replaced with a semispace in calculations of the Casimir
pressure \cite{32}.  For a freestanding in a vacuum peptide film, one should
put $\eps^{(2)}(\omega)=1$. The Casimir pressure of the peptide coating (film)
at temperature $T$ is given by the Lifshitz theory
\begin{eqnarray}
&&
P(a)=-\frac{k_BT}{\pi}\sum_{l=0}^{\infty}{\vphantom{\sum}}^{\prime}
\int_0^{\infty}\!\!k p^{(1)}\kk dk
\label{eq1} \\
&&~~\times\sum_{\alpha}
\frac{1}{\left[r_{\alpha}^{(1,0)}\kk r_{\alpha}^{(1,2)}\kk\right]^{-1}
e^{2ap^{(1)}\kk }-1}.
\nonumber
\end{eqnarray}
\noindent
Here,  $k_B$ is the Boltzmann constant, $\xi_l=2\pi k_BTl/\hbar$ with
$l=0,\,1,\,2,\,\ldots$ are the Matsubara frequencies, the prime on the
sum in $l$ divides the term with $l=0$ by two,
$k$ is the magnitude of the wave vector
projection on the plane of peptide film,
the sum in $\alpha$ is  over
two independent polarizations of the electromagnetic field,
transverse electric  ($\alpha={\rm TE}$) and transverse magnetic
($\alpha={\rm TM}$), and
\begin{equation}
p^{(n)}\kk=\sqrt{k^2+\eps^{(n)}(\il)\mu^{(n)}(\il)
\frac{\xi_l^2}{c^2}},
\label{eq2}
\end{equation}
\noindent
where $n=0,\,1,\,2$.

Equation (\ref{eq1}) also contains the reflection coefficients on the
boundary planes between a film and a vacuum
\begin{eqnarray}
&&
r_{\rm TM}^{(1,0)}\kk=\frac{p^{(1)}\kk-\eps^{(1)}(\il)
p^{(0)}\kk} {p^{(1)}\kk+\eps^{(1)}(\il)p^{(0)}\kk},
\nonumber \\[-1.5mm]
&&\label{eq3}\\[-1.5mm]
&&
r_{\rm TE}^{(1,0)}\kk=\frac{p^{(1)}\kk-\mu^{(1)}(\il)
p^{(0)}\kk} {p^{(1)}\kk+\mu^{(1)}(\il)p^{(0)}\kk},
\nonumber
\end{eqnarray}
\noindent
and between a film and a substrate
\begin{eqnarray}
&&
r_{\rm TM}^{(1,2)}\kk=\frac{\eps^{(2)}(\il)p^{(1)}\kk-\eps^{(1)}(\il)
p^{(2)}\kk} {\eps^{(2)}(\il)p^{(1)}\kk+\eps^{(1)}(\il)p^{(2)}\kk},
\nonumber \\
&&
r_{\rm TE}^{(1,2)}\kk=\frac{p^{(1)}\kk-\mu^{(1)}(\il)
p^{(2)}\kk} {p^{(1)}\kk+\mu^{(1)}(\il)p^{(2)}\kk}.
\label{eq4}
\end{eqnarray}

Equation (\ref{eq1}) has been extensively used to investigate the Casimir
effect in inorganic (metallic and dielectric) films and coatings
\cite{33,34,35,36,37,38}. It allows computation of the Casimir pressure
as a function of film thickness by the known quantities
$\eps^{(n)}(\il)$ and $\mu^{(1)}(\il)$. In doing so, at room temperature
the result depends only on $\mu^{(1)}(0)$ \cite{39}.

An application of the same approach to peptide films is not a simple task.
The point is that for typical peptides the optical data over the wide
frequency range are not available. What is more, peptide films usually
contain some volume fraction of a plasticizer whose role may be played
by water \cite{40,41}. If the peptide film is doped with nanoparticles,
this should be taken into account
in its effective dielectric permittivity.

In Ref.~\cite{29} the dielectric permittivity of a model peptide
$\eps^{(p)}(\il)$ along the imaginary frequency axis was composed from the
imaginary parts of the permittivities of electrically neutral 18-residue zinc
finger peptide in the microwave region \cite{42} and of cyclic tripeptide
RGD-4C in the region of ultraviolet frequencies \cite{43}.
The permittivity of water, $\eps^{(w)}(\il)$, was used in the
representation of Ref.~\cite{44}. The molecules of the model peptide
 are assumed to have an irregular shape, be a few nanometers in
size and randomly distributed in water. Under an assumption that a peptide
film contains the volume fraction of water $\Phi$, the film dielectric
permittivity, $\eps_{\Phi}^{(p)}(\il)$, is then found from the mixing
formula \cite{45}
\begin{equation}
\frac{\eps_{\Phi}^{(p)}(\il)-1}{\eps_{\Phi}^{(p)}(\il)+2}=
\Phi\frac{\eps^{(w)}(\il)-1}{\eps^{(w)}(\il)+2}+
(1-\Phi)\frac{\eps^{(p)}(\il)-1}{\eps^{(p)}(\il)+2},
\label{eq5}
\end{equation}
\noindent
which is a consequence of the Clausius-Mossotti equation.
The obtained dielectric permittivities $\eps_{\Phi}^{(p)}(\il)$ for
$\Phi=0$, 0.1, 0.25, and 0.4 as the functions of $\xi$ are presented in
Fig.~2 of Ref.~\cite{29}. It is seen that
$\eps_{0}^{(p)}(\il)=\eps^{(p)}(\il)$.

\begin{figure}[b]
\vspace*{-1.5cm}
\centerline{\hspace*{1cm}
\includegraphics[width=14 cm]{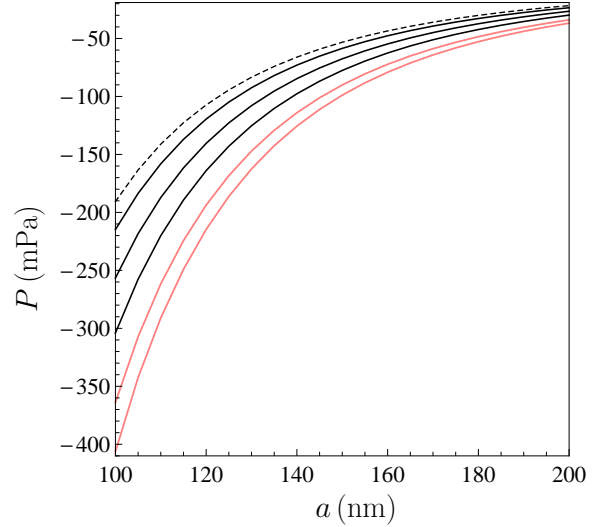}}
\vspace*{-11.7cm}
\caption{\label{fig1} The Casimir pressures in freestanding undoped peptide films
(the dashed line is for a pure peptide and the following three black
solid lines from top to bottom are for the films containing $\Phi = 0.1$,
 0.25, and 0.4 fractions of water, respectively) and in doped
with Au nanoparticles peptide films with $\Phi = 0.4$  (the lowest
and the next to it solid lines are for the films containing $\beta =0.05$
 and 0.03 volume fractions of nanoparticles, respectively) are
shown as the functions of film thickness at $T = 300~$K.}
\end{figure}
Substituting $\eps^{(1)}=\eps_{\Phi}^{(p)}$ and
$\mu^{(1)}=\eps^{(2)}=1$ in Eq.~(\ref{eq1}), we have computed the
Casimir pressure in the freestanding in a vacuum undoped peptide film
containing different volume fractions $\Phi$ of water at $T=300~$K.
The computational results are shown in Fig.~\ref{fig1} by the dashed line ($\Phi=0$)
and by the following three black solid lines counted from top to bottom
for $\Phi=0.1$, 0.25, and 0.4, respectively, as the functions of film
thickness. As is seen in Fig.~\ref{fig1}, the pressure  is
negative which corresponds to an attraction and makes film more stable.

Now we consider the peptide film containing the volume fraction $\Phi$ of water
and assume that it is then doped with Au nanoparticles of spherical shape occupying
the volume fraction $\beta$ of the obtained film. The dielectric permittivity
of Au nanoparticles along the imaginary frequency axis $\eps_{\rm Au}(\il)$ is
found using the optical data for Au \cite{46} and was extensively used in
calculations of the Casimir force \cite{21,22,23}. As a result, the dielectric
permittivity of doped peptide film is given by the Maxwell-Garnet mixing
formula \cite{47}
\begin{equation}
\eps_{\Phi,\beta}^{(p)}=\eps_{\Phi}^{(p)}
\left(1+\frac{3\beta X}{1-\beta X}\right),\quad
X=\frac{\eps_{\rm Au}-\eps_{\Phi}^{(p)}}{\eps_{\rm Au}+2\eps_{\Phi}^{(p)}}.
\label{eq6}
\end{equation}

Computations of the Casimir pressure are made at $T=300~$K for the doped
freestanding peptide films using Eq.~(\ref{eq1}) where
$\eps^{(1)}=\eps_{\Phi,\beta}^{(p)}$,
$\mu^{(1)}=\eps^{(2)}=1$, $\Phi=0.4$, and $\beta=0.03$ or 0.05.
The computational results are shown in Fig.~\ref{fig1} as the functions of
film thickness by the lowest and next to it solid lines ($\beta=0.05$ and 0.03,
respectively). It is seen that these lines demonstrate much larger in
magnitude Casimir pressures than the bottom black line which holds for
 an undoped film with $\Phi=0.4$.

\begin{figure}[b]
\vspace*{-0.7cm}
\centerline{\hspace*{1cm}
\includegraphics[width=14 cm]{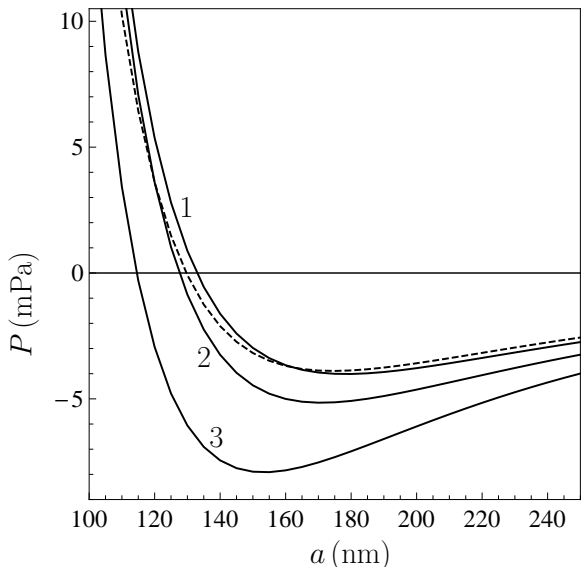}}
\vspace*{-11.7cm}
\caption{\label{fig2} The  Casimir pressures
in undoped peptide films deposited on a
SiO$_2$ substrate are shown as the dashed line for a pure peptide and
by the lines 1, 2, and 3 for the films containing $\Phi = 0.1$, 0.25,
and 0.4 fractions of water, respectively, as the functions of film
thickness at $T = 300~$K.}
\end{figure}
Now  we turn our attention to the most interesting case from the practical
standpoint, i.e., to peptide films deposited on a dielectric substrate.
As a substrate material we use SiO$_2$ glass \cite{6} which dielectric
permittivity $\eps^{(2)}(\ri\xi)$ has an accurate analytic
representation \cite{45}. First we calculate the Casimir pressure in an
undoped coating using Eq.~(\ref{eq1}) where
$\eps^{(1)}=\eps_{\Phi}^{(p)}$ and $\mu^{(1)}=1$.
The computational results as functions of the film thickness are shown
in Fig.~\ref{fig2} by the dashed line for a pure peptide coating and
by the lines 1, 2, and 3 for peptide coatings containing the fractions
$\Phi=0.1$, 0.25, and 0.4 of water, respectively.
As is seen in Fig.~\ref{fig2}, for pure peptide coating of less than 130~nm
thickness the fluctuation-induced Casimir pressure becomes positive which
makes the film less stable. The same holds for coatings containing
$\Phi=0.1$, 0.25, and 0.4 fractions of water if they are thinner than
133, 128, and 115~nm, respectively. The maximum in magnitude negative
Casimir pressures contributing to the coating stability are reached for
the film thicknesses $a=175$, 180, 170, and 155~nm for the fractions of
water in the film $\Phi=0$, 0.1, 0.25, and 0.4, respectively.

\begin{figure}[b]
\vspace*{-3.8cm}
\centerline{\hspace*{1cm}
\includegraphics[width=14 cm]{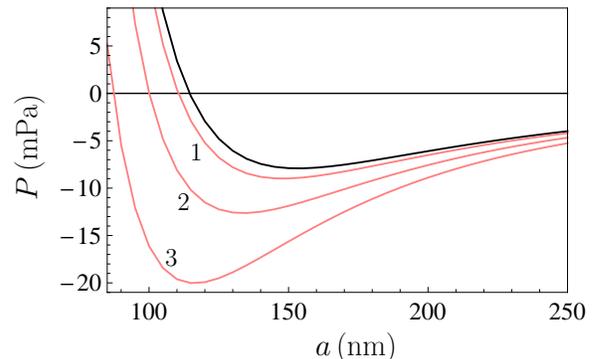}}
\vspace*{-11.7cm}
\caption{\label{fig3} The  Casimir pressures
in doped with Au nanoparticles peptide
films with $\Phi = 0.4$ fraction of water deposited on a SiO$_2$
substrate are shown by the lines 1, 2, and 3 for the volume fractions
of nanoparticles $\beta =0.01$, 0.03, and 0.05, respectively, as the
functions of film thickness at  $T = 300~$K. The black line with no
number presents similar results for an undoped coating. }
\end{figure}

Next we consider the peptide coating doped with Au nanoparticles. As in the case
of a freestanding film, the $\Phi=0.4$ fraction of water in the film
before doping is assumed. The computations are again performed by Eq.~(\ref{eq1})
where now $\eps^{(1)}=\eps_{\Phi,\beta}^{(p)}$, $\mu^{(1)}=1$,
and $\eps^{(2)}$ is the dielectric permittivity of a SiO$_2$ substrate.
The computational results for the Casimir pressure in peptide coatings
are shown in Fig.~\ref{fig3} as functions of the film thickness by the lines 1, 2,
and 3 for the doped films with  the volume fractions of Au nanoparticles
$\beta=0.01$, 0.03, and 0.05, respectively.
For comparison purposes, the black line with no number reproduces the line 3
in Fig.~\ref{fig2} showing the Casimir pressure in an undoped peptide coating
containing $\Phi=0.4$ fraction of water.

As is seen in Fig.~\ref{fig3}, the presence of doping widens the range of film
thicknesses where the Casimir pressure is negative and makes the minima deeper.
Specifically, for the fractions $\beta=0.01$, 0.03, and 0.05 of nanoparticles
in the peptide coating the pressure changes its sign from negative to positive
for film thickness below 110, 100, and 87~nm, respectively, to compare with
115~nm for an undoped coating (the black line with no number in Fig.~\ref{fig3}).
The largest magnitudes of the negative Casimir pressure at the points of minimum
are 8.98, 12.63, and 20.0~mPa reached for the film thicknesses
$a=150$, 135, and 115~nm, respectively (compared to 7.91~mPa reached at $a=155~$nm
for an undoped peptide coating). Thus, an addition of Au nanoparticles makes
the peptide coating more stable over a wider range of film thicknesses.

Finally we consider the case of peptide coatings doped with magnetic nanoparticles
which endows the coating with superparamagnetic properties. The computations
below are performed for the iron oxide (Fe$_3$O$_4$) magnetite nanoparticles which
are often used in ferrofluids \cite{48,49}. The dielectric permittivity of
magnetite along the imaginary frequency axis, $\eps_m(\i\xi)$, was found in
Ref.~\cite{49} basing on the measurement data of Ref.~\cite{50}.
Then, the dielectric permittivity $\eps_{\Phi,\beta}^{(p)}(\il)$ of peptide
film containing some fraction of water and doped with magnetite nanoparticles
was calculated by Eq.~(\ref{eq6}) where $\eps_{\rm Au}$ was replaced
with $\eps_m$. It was used as $\eps^{(1)}(\il)$ in Eq.~(\ref{eq1}).

The magnetic permeability of peptide coating doped with magnetic nanoparticles
deserves special attention. As was mentioned above, the magnetic properties
influence the Casimir pressure only through the term of Eq.~(\ref{eq1})
with $l=0$. This is explained by the fact that at $T=300~$K the magnetic
permeability drops to unity at much smaller frequencies than $\xi_1$
\cite{39}. For the static magnetic permeability of peptide coating which
is a superparamagnetic system containing the fraction $\beta$ of single-domain
magnetic nanoparticles of radius $R$ one obtains \cite{51}
\begin{equation}
\mu^{(1)}(0)=1+\frac{16\pi^2R^3\beta M_s^2}{9k_BT},
\label{eq7}
\end{equation}
\noindent
where $M_s\approx 300~\mbox{emu/cm}^3=3\times 10^5~$A/m is the saturation
magnetization per unit volume for a single nanoparticle of magnetite \cite{52}.

We note that according to this equation the magnetic permeability of doped peptide
film depends not only on the volume fraction of nanoparticles $\beta$ but also on
their radius. This is not the case for the dielectric permittivity of the same
coating which, according to Eq.~({\ref{eq6}), is completely determined by the
value of $\beta$. We use the typical value $R=5~$nm in below computations.
Then, for $\beta=0.01$, 0.03, and 0.05 one obtains from Eq.~(\ref{eq7}) for
the static magnetic permeability $\mu^{(1)}(0)=1.05$, 1.14, and 1.24,
respectively.

\begin{figure}[t]
\vspace*{-1.cm}
\centerline{\hspace*{1cm}
\includegraphics[width=14 cm]{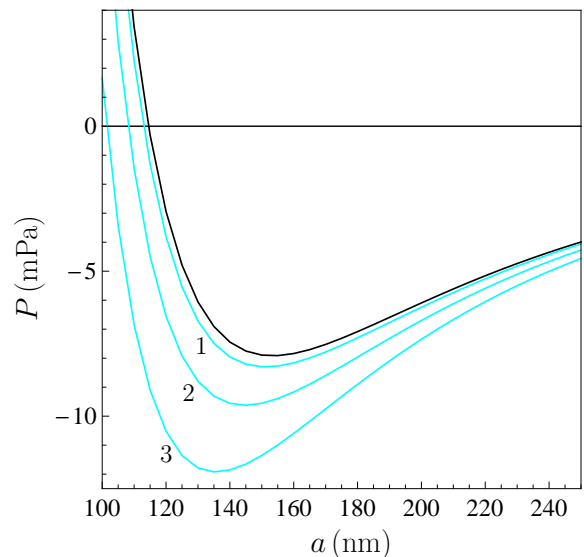}}
\vspace*{-11.7cm}
\caption{\label{fig4} The  Casimir pressures
in doped with magnetite nanoparticles peptide films
containing $\Phi = 0.4$ fraction of water deposited on a SiO$_2$
substrate are shown by the lines 1, 2, and 3 for the volume fractions
of nanoparticles $\beta =0.01$, 0.03, and 0.05, respectively,  as the
functions of film thickness at  $T = 300~$K. The black line with no
number presents similar results for an undoped coating. }
\end{figure}
Computations of the Casimir pressure in peptide coatings doped with magnetite
nanoparticles were performed by Eq.~(\ref{eq1}) using the dielectric permittivity
and magnetic permeability obtained above as well as the  dielectric permittivity
of a SiO$_2$ substrate $\eps^{(2)}(\il)$. In so doing the $\Phi=0.4$ volume
fraction of water in the film was assumed. The computational results are
shown in Fig.~\ref{fig4} by the lines 1, 2, and 3 as functions of the film
thickness for the volume fractions of nanoparticles $\beta=0.01$, 0.03, and 0.05,
respectively, at $T=300~$K. The black line with no number shows the Casimir
pressure in an undoped peptide coating with the same fraction of water.
As is seen in Fig.~\ref{fig4}, the presence of magnetite nanoparticles in
peptide coating again widens the range of film thicknesses where the pressure
is negative and  makes the minima deeper. Thus, the pressure changes its
sign at $a=113$, 108, and 102~nm for peptide coatings with $\beta=0.01$, 0.03,
and 0.05, respectively (to compare with $a=115~$nm for an undoped film).
The maximum magnitudes of the Casimir pressure 8.3, 9.6, and 11.9~mPa are
reached at the extremum points of 150, 145 and 135~nm for $\beta=0.01$, 0.03, and
0.05, respectively. This should be compared with 7.91~mPa at $a=155~$nm for an
undoped coating. Thus, doping with magnetic nanoparticles results in more stable
peptide films possessing superparamagnetic properties.

In the foregoing, we have calculated the fluctuation-induced (Casimir)
pressure in thin peptide films and coatings, both undoped and doped with
metallic nanoparticles, which makes an impact on the film stability.
Estimations show that for films of about 100 nm thickness the
electromagnetic fluctuations may contribute up to 20\% of the total
cohesive energy of a film \cite{29,53} and all the more for thinner
films. This adds in importance to the effect of increased
stability of peptide films and coatings due to their doping with
metallic nanoparticles. Taking into account that stability is the crucial
property of a coating, the harnessing of doped peptide films may become
beneficial in various applications mentioned above. The proposed doping
of a peptide coating with magnetic nanoparticles not only increases the
film stability, but makes it superparamagnetic which opens further
application areas for bioelectronics, such as in spintronics and magnetic
resonance imaging.

The authors were supported by the Peter the Great Saint
Petersburg Polytechnic University in the framework of the Program
``5-100-2020". The work of V.M.M. was partially funded by the
Russian Foundation for Basic Research, Grant No. 19-02-00453 A.
His work was also partially supported by the Russian Government
Program of Competitive Growth of Kazan Federal University.



\begin{thebibliography}{99}
\bibitem{1}
A.~Ulman,
{\it An Introduction to Ultrathin Organic Films: From Langmuir-Blodgett
to Self-Assembly}
(Academic Press, London, 1991).
\bibitem{2}
G.~Meller and T.~Grasser (eds.),
{\it Organic Electronics} (Springer, Heidelberg, 2010).
\bibitem{3}
Chun-Yi Lee, Jenn-Chang Hwang, Yu-Lun Chueh, Ting-Hao Chang, Yi-Yun Cheng,
and Ping-Chiang Lyu,
Hydrated bovine serum albumin as the gate dielectric material for organic
field-effect transistors,
Org. Electr. {\bf 14}, 2645 (2013).
\bibitem{4}
Mingchao Ma, Xinjun Xu, Leilei Shi, and Lidong Li,
Organic field-effect transistors with a low driving voltage using albumin
as the dielectric layer,
RSC Advances {\bf 4}, 58720 (2014).
\bibitem{5}
C.~D.~Dimitrakopoulos and P.~R.~L.~Malenfant,
Organic thin film transistors for large area electronics,
Adv. Mater. {\bf 14}, 99 (2002).
\bibitem{6}
M.~Righi, G.~L.~Puleo, I.~Tonazzini, G.~Giudetti, M.~Cecchini, and S.~Micera,
Peptide-based coatings for flexible implantable neural interfaces,
Sci. Reports {\bf 8}, 502 (2018).
\bibitem{7}
T.~Guterman and E.~Gazit,
Toward peptide-based bioelectronics: reductionist design of conductive
pili mimetics,
Bioelectron. Med. (Lond.) {\bf 1}, 131 (2018).
\bibitem{8}
J.~Yu, J.~R.~Horsley, and A.~D.~Abell,
Peptides as Bio-Inspired Electronic Materials: An Electrochemical and
First-Principles Perspective,
Acc. Chem. Res. {\bf 51}, 2237 (2018).
\bibitem{9}
S.~S.~Panda, H.~E.~Katz, and J.~D.~Tovar,
Solid-state electrical applications of protein and peptide based nanomaterials,
Chem. Soc. Rev. {\bf 47}, 3640 (2018).
\bibitem{10}
B.~Li, D.~T.~Haynie, N.~Palath, and D.~Janisch,
Nano-Scale Biomimetics: Fabrication and Optimization of Stability of
Peptide-Based Thin Films,
J. Nanosci. Nanotech. {\bf 5}, 2042 (2005).
\bibitem{11}
P.~Fattahi, G.~Yang, G.~Kim, and M.~R.~Abidian,
A Review of Organic and Inorganic Biomaterials for Neural Interfaces,
Adv. Mater. {\bf 26}, 1846 (2014).
\bibitem{12}
A.~Arul, S.~Sivagnanam, A.~Dey, O.~Mukherjee, S.~Ghosh, and P.~Das,
The design and development of short peptide-based novel smart materials to
prevent fouling by the formation of non-toxic and biocompatible coatings,
RSC Advances {\bf 10}, 13420 (2020).
\bibitem{13}
S.~Sharma, R.~W.~Johnson, and T.~A.~Desai,
Evaluation of the stability of nonfouling ultrathin poly(elhylen glucol)
films for silicon-based microdevices,
Langmuir {\bf 20}, 348 (2004).
\bibitem{14}
M.~Natesan and R.~G.~Ulrich,
Protein microarrays, and biomarkers of infection disease,
Int. J. Mol. Sci. {\bf 11}, 5165 (2010).
\bibitem{15}
C.-K.~Chou, N.~Jing, H.~Yamaguchi, P.-H.~Tsou, H.-H.~Lee, C.-T.~Chen,
Y.-N.~Wang, S.~Hong, C.~Su, J.~Kameoka, and M.-C.~Hung,
Rapid detection of two-protein interaction with a single fluorophore by
using a microfluidic device,
Analyst {\bf 135}, 2907 (2010).
\bibitem{16}
H.~Chandra, P.~J.~Reddy, and S.~Srivastava,
Protein microarrays and novel detection platforms,
Expert Rev. Proteomics {\bf 8}, 61 (2011).
\bibitem{17}
Pui Mun Lee, Ze Xiong, and John Ho,
Methods for powering bioelectronic microdevices,
Bioelectron. Med. (Lond.) {\bf 1}, 201 (2018).
\bibitem{18}
M.~Kardar and R.~Golestanian,
The ``friction" of vacuum, and other fluctuation-induced forces,
Rev. Mod. Phys. {\bf 71}, 1233 (1999).
\bibitem{19}
R.~H.~French, V.~A.~Parsegian, R.~Podgornik et al.,
Long range interactions in nanoscale science,
Rev. Mod. Phys. {\bf 82}, 1887 (2010).
\bibitem{20}
V.~A.~Parsegian,
{\it Van der Waals Forces: A Handbook for Biologists, Chemists, Engineers,
and Physicists}
(Cambridge University Press, Cambridge, 2005).
\bibitem{21}
M.~Bordag, G.~L.~Klimchitskaya, U.~Mohideen, and V.\ M.\ Mostepanenko,
{\it Advances in the Casimir Effect}
(Oxford University Press, Oxford, 2015).
\bibitem{22}
G.~L.~Klimchitskaya, U.~Mohideen, and V.~M.~Mostepanenko,
The Casimir force between real materials: Experiment and theory,
Rev. Mod. Phys. {\bf 81}, 1827 (2009).
\bibitem{23}
L.~M.~Woods, D.~A.~R.~Dalvit, A.~Tkatchenko, P.~Rodriguez-Lopez,
A.~W.~Rodriguez, and R.~Podgornik,
Materials perspective on Casimir and van der Waals interactions,
Rev. Mod. Phys. {\bf 88}, 045003 (2016).
\bibitem{24}
E.~M.~Lifshitz and L.~P.~Pitaevskii,
{\it Statistical Physics, Pt. II}
(Pergamon Press, Oxford, 1980).
\bibitem{25}
V.~A.~Parsegian and B.~W.~Ninham,
Application of the Lifshitz theory to the calculation of
van der Waals forces across thin lipid films,
Nature {\bf 224}, 1197 (1972).
\bibitem{26}
S.~Nir,
Van der Waals interactions between surfaces of biological interest,
Progr. Surf. Sci. {\bf 8}, 1 (1976).
\bibitem{27}
C.~M.~Roth, B.~L.~Neal, and A.~M.~Lenhoff,
Van der Waals interactions involving proteins,
Biophys. J. {\bf 70}, 977 (1996).
\bibitem{28}
Bing-Sui Lu and R. Podgornik,
Effective interactions between fluid membranes,
Phys. Rev. E {\bf 92}, 022112 (2015).
\bibitem{29}
M.~A.~Baranov, G.~L.~Klimchitskaya, V.~M.~Mostepanenko, and E.~N.~Velichko,
Fluctuation-induced free energy of thin peptide films,
Phys. Rev. E {\bf 99}, 022410 (2019).
\bibitem{30}
E.~N.~Velichko, M.~A.~Baranov, and V.~M.~Mostepanenko,
Change of sign in the Casimir interaction of peptide films deposited on a
dielectric substrate,
Mod. Phys. Lett. A {\bf 35}, 2040020 (2020).
\bibitem{31}
Jinmao Yan, Yunxiang Pan, A.~G.~Cheetham, Yi-An Lin, Wei Wang, Honggang Cui,
and Chang-Jun Liu,
One-Step Fabrication of Self-Assembled Peptide Thin Films with Hightly Dispersed
Noble Metal Nanoparticles,
Langmuir {\bf 29}, 16051 (2013).
\bibitem{32}
G.~L.~Klimchitskaya and V.~M.~Mostepanenko,
Observability of thermal effects in the Casimir interaction with graphene-coated
substrates,
Phys. Rev. A {\bf 89}, 052512 (2014).
\bibitem{33}
G.~L.~Klimchitskaya and V.~M. ~Mostepanenko,
Casimir free energy of metallic films: Discriminating between Drude and plasma
model approaches,
Phys. Rev. A {\bf 92}, 042109 (2015).
\bibitem{34}
G.~L.~Klimchitskaya and V.~M.~Mostepanenko,
Casimir and van der Waals energy of anisotropic atomically thin metallic films,
Phys. Rev. B {\bf 92}, 205410 (2015).
\bibitem{35}
G.~L.~Klimchitskaya and V.~M.~Mostepanenko,
Casimir free energy and pressure for magnetic metal films,
Phys. Rev. B {\bf 94}, 045404 (2016).
\bibitem{36}
G.~L.~Klimchitskaya and V.~M.~Mostepanenko,
Characteristic properties of the Casimir free energy for metal films deposited
on metallic plates,
Phys. Rev. A {\bf 93}, 042508 (2016).
\bibitem{37}
G.~L.~Klimchitskaya and V.~M.~Mostepanenko,
Low-temperature behavior of the Casimir free energy and entropy of metallic
films,
Phys. Rev. A {\bf 95}, 012130 (2017).
\bibitem{38}
G.~L.~Klimchitskaya and V.~M.~Mostepanenko,
Casimir free energy of dielectric films: Classical limit, low-temperature
behavior and control,
J. Phys.: Condens. Matter {\bf 29}, 275701 (2017).
\bibitem{39}
B.~Geyer, G.~L.~Klimchitskaya, and V.~M.~Mostepanenko,
Thermal Casimir interaction between two magnetodielectric plates,
Phys. Rev. B {\bf 81}, 104101 (2010).
\bibitem{40}
A. Gennadios (ed.),
{\it Protein-Based Films and Coatings}
(CRC Press, Boca Raton, 2002).
\bibitem{41}
N.~Gontard and S.~Ring,
Edible wheat gluten film: Influence of water content on glass transition
temperature,
J. Agric. Food Chem. {\bf 44}, 3474 (1996).
\bibitem{42}
G.~L\"{o}ffler, H.~Schreiber, and O.~Steinhauser,
Calculation of the Dielectric Properties of a Protein and its Solvent:
Theory and a Case Study,
J. Mol. Biol. {\bf 270}, 520 (1997).
\bibitem{43}
P.~Adhikari, A.~M.~Wen, R.~H.~French, V.~A.~Parsegian, N.~F.~Steinmetz,
R.~Podgornik, and W.-Y.~Ching,
Electronic Structure, Dielectric Response, and Surface Charge Distribution of
RGD (1FUV) Peptide,
Sci. Reports. {\bf 4}, 5605 (2014).
\bibitem{44}
L.~Bergstr\"{o}m,
Hamaker constant of inorganic materials,
Adv. Coll. Interface Sci. {\bf 70}, 125 (1997).
\bibitem{45}
D.~B.~Hough and L.~H.~White,
The calculation of Hamaker constant from Lifshitz theory with
application to wetting phenomena,
Adv. Coll. Interface Sci. {\bf 14}, 3 (1980).
\bibitem {46}
{\it Handbook of Optical Constants of Solids},
ed. E.~D.~Palik (Academic, New York, 1985).
\bibitem{47}
A.~H.~Sihvola,
{\it Electromagnetic Mixing Formulas and Applications}
(The Institution of Electrical Engineers, London, 1999).
\bibitem{48}
T.~Guo, X.~Bian, and C.~Yang,
A new method to prepare water based Fe$_3$O$_4$ ferrofluid with high
stabilization,
Physica A: Stat. Mech. Applic. {\bf 438}, 560 (2015).
\bibitem{49}
G.~L.~Klimchitskaya, V.~M.~Mostepanenko, E.~K.~Nepomnyashchaya, and
E.~N.~Velichko,
Impact of magnetic nanoparticles on the Casimir pressure in three-layer
systems,
Phys. Rev. B {\bf 99}, 045433 (2019).
\bibitem{50}
A.~Schlegel, S.~F.~Alvarado, and P.~Wachter,
Optical properties of magnetite (Fe$_3$O$_4$),
J. Phys. C: Solid State Phys. {\bf 12}, 1157 (1979).
\bibitem{51}
S.~V.~Vonsovskii,
{\it Magnetism}
(Wiley, New York, 1974).
\bibitem{52}
S.~van Berkum, J.~T.~Dee, A.~P.~Philipse, and B.~E.~Ern\'{e},
Frequency-dependent magnetic susceptibility of magnetite and cobalt
ferrite nanoparticles embedded in PAA hydrogel,
Int. J. Mol. Sci. {\bf 14}, 10162 (2013).
\bibitem{53}
I.~Boinovich and A.~Emelyanenko,
Wetting and surface forces,
Adv. Coll. Interface Sci. {\bf 165}, 60 (2011).
\end{thebibliography}
\end{document}